# Theoretical development in the viscosity of ferrofluid

Anupam Bhandari

Department of Mathematics, School of Engineering

University of Petroleum & Energy Studies (UPES)

Energy Acres Building, Bidholi

Dehradun- 248007, Uttarakhand, India

E-mail: pankaj.anupam6@gmail.com

**https://www.upes.ac.in**

**Abstract:**

The viscosity of ferrofluid has an important role in liquid sealing of the hard disk drives, biomedical applications as drug delivery, hyperthermia, and magnetic resonance imaging. In the absence of a magnetic field, the viscosity of ferrofluid depends on the volume concentration of magnetic nanoparticles including surfactant layers. However, under the influence of a stationary magnetic field, the viscosity of ferrofluid depends on the angle between the applied magnetic field and vorticity in the flow. If this angle is 90$^\circ$, then there is a maximum increase in the viscosity. If the magnetic field and the vorticity in the flow are parallel to each other, then there is no change in the viscosity since the applied magnetic field does not change the speed of the rotation of magnetic nanoparticles in the fluid.

The viscosity of ferrofluid in the presence of an alternating magnetic field demonstrates interesting behavior. When field frequency matches with the relaxation time, known as resonance condition, then there is no impact of an alternating magnetic field in the viscosity of ferrofluid. If the frequency of an alternating magnetic field is less than resonance frequency, then an alternating magnetic field increases the viscosity of ferrofluid. Using higher frequency than resonance condition reduces the viscosity of ferrofluid and researchers reported this incident as the negative viscosity effect. If the frequency of an alternating magnetic field tends to infinite, then ferrofluid ceases to feel a magnetic field. In this case, there is no impact of an alternating magnetic field on the viscosity of ferrofluid.

**Keywords:** Ferrofluids, viscosity, magnetic field, volume concentration, negative viscosity, ferrohydrodynamics.

# 1. Introduction to Ferrofluids

*1.1. Formation of ferrofluid*

Ferrofluid is also known as magnetic fluid. Ferrofluids are not directly available in nature[1]. These fluids are artificially synthesized of colloidal mixtures of carrier liquid, typically water or oil, and magnetic nanoparticles[1], [2]. Surfactants are used in the colloidal mixtures to ensure the stability of the ferrofluid[1], [3]. Surfactants prevent the agglomeration of the magnetic particles[4], [5]. Ferrofluids, which work at zero gravity region, are recognized as colloidal suspension of superparamagnetic materials[1], [4], [5]. In ferrofluid preparation, we consider the size of the magnetic particles 5-15 nanometers in diameter and volume fraction up to approximately 10%[1], [3]. We select water-based carrier liquid for medical purposes, mineral oil and silicon organic-based carrier liquid for lubrication and sealing system, and hydrocarbon-based carrier liquids for printing devices[4], [5]. Properties of the ferrofluid depend on the size of the magnetic particles and their magnetization[1], [3]. The stability of ferrofluid is ensured by the thermal motion which prevents the agglomeration and precipitation[4], [5]. Thermal motion increases with decreasing the size of the particles. The magnetic properties disappear if the size of magnetic particles is less than 1-2nanometers[4], [5]. Surfactant is important for the stability of ferrofluid and long-chain molecules (e.g., $OOH$, $H_2OH$, $H_2NH_2$, and so on) are used for surfactants[4], [5]. Surfactants produce the chemical reaction in the colloidal mixture and this reaction reduces the size of the magnetic particles[5]. Reduction of the size of magnetic particles loses magnetic properties[5].

For the application of ferrofluid, it is essentially required that ferrofluid should be very stable concerning the temperature in the presence of a magnetic field[6]. Therefore, agglomeration of the magnetic particles must be avoided for proper commercial use[6]. Kilkuchi et al. described experimentally that the reaction temperature from 200 to 250 $^oC$, the size of the magnetic particles increases from 5 to 11 nanometers[7]. Considering the nonhomogeneous distribution of pH and dielectric constant, the microemulsion method is useful to prepare stable ferrofluid[8]. Imran et al. synthesized highly stable ferrofluid using motor oil as base fluid and found 13 nanometers average particle size of γ-Fe2O3[9]. The stability of ferrofluid remains protective for small particles size since their concentrations and dipolar coupling energies are too low for field-induced dipolar structure formation[10]. D. P. Lalas and S. Carmi investigated the stability of motionless ferrofluid using the concept of Rayleigh number[11]. Tari et al. have investigated the role of magnetization and temperature for the stability of diester-based $Fe_3O_4$ ferrofluid[12]. In the transition from the laminar to turbulence motion of ferrofluid, we need to check the accuracy of the numerical solution along with the stability and uniqueness[13]. Thermal and magnetic stress is an important class of surface interactions of magnetic particles near Curie temperature[14].

Stability of ferrofluid in the presence of non-uniform magnetic field investigated through stability coefficients using microsensors[15]. Internal structures and macroscopic physical properties[16] and arrangements of the magnetic particles in special structures[17] can serve in the development of the applications of the magnetic fluid. There is no suitable procedure to explain the thermodynamical and dynamical properties of magnetic fluids with developed microstructure[18]. There is a critical chain number for phase transition in ferrofluid[19] and the magnetic field exhilarates the formation of chains from Ferro-particles[20]. Wiedenmann investigated the stability of nanoparticles in ferrofluid against electrostatic repulsion or surfactants[21]. Gazeau et al. demonstrated the Brownian motion nanoparticles in ferrofluid under applied magnetic field[22]. Sousa et al. investigated the surface magnetic properties of $NiFe_2O_4$ nanoparticles[23]. Raikher et al. demonstrated a magneto-optical way to analyze the internal and external magnetic relaxation in magnetic fluids[24]. Raikher et al. explained the particle orientation dynamics using the general Fokker-Planck equation[25]. The synthesis of ferrofluid, stability, and characteristics with magnetic properties have been investigated for different types of magnetic nanoparticles[26]–[28].

*1.2. Real-life applications of ferrofluids*

In the mid-1960s, ferrofluid was developed by NASA as a method for controlling fluid in space since the flow of ferrofluid can be controlled by the external magnetic field. Conventional ferrofluids are useful in liquid seals, shock absorbers, controlling heat in loudspeakers, printing for paper money, and lubrication bearings[1]–[5]. The commercial use of ferrofluids has been available in the literature[29], [30]. Ferrofluids can be used in heat transfer and damping problems[31]. The viscosity of ferrofluid has an important role in these applications[32]. Ferrofluids have an important role in biomedical applications for diagnostic and therapy. Drug delivery, hyperthermia treatments, and magnetic resonance imaging are some of the major applications in the field of biomedical engineering[33]–[35]. Fundamental and applied research in the ferrofluid, nowadays, researchers are trying to develop magneto-optical devices and ferrofluidic sensors[36], [37]. Even, researchers have shown the application of ferrofluid in environmental engineering[38].

## 2. Governing equations in ferrohydrodynamic flow

To describe the behavior of ferrofluid in different flow domains, the researchers use the following set of equations[39]–[42]:

The equation of continuity[1], [43], [44]

$$\nabla \cdot v = 0 \qquad (1)$$

The equation of motion[39], [44]–[46]

$$\rho \frac{dv}{dt} = -\nabla p + \mu \nabla^2 v + \mu_0 (M \cdot \nabla) H + \frac{1}{2\tau_s} \nabla \times (\omega_p - \Omega) \qquad (2)$$

The equation of magnetization[3], [40]

$$\frac{dM}{dt} = \boldsymbol{\omega}_{\mathrm{p}} \times \boldsymbol{M} - \frac{1}{\tau_B}(\boldsymbol{M} - \boldsymbol{M_0}) \qquad (3)$$

The equation of rotational motion[39], [44], [47]

$$I\frac{d\boldsymbol{\omega}_{\mathrm{p}}}{dt} = \boldsymbol{M} \times \boldsymbol{H} - \frac{I}{\tau_s}(\boldsymbol{\omega}_{\mathrm{p}} - \boldsymbol{\Omega}) \qquad (4)$$

The equation of instantaneous magnetization[4][1]–[3], [5]

$$\boldsymbol{M_0} = nmL(\xi)\frac{H}{H}, \qquad \xi = \frac{mH(t)}{k_B T}, \qquad L(\xi) = \coth\xi - \xi^{-1} \qquad (5)$$

The Energy Equation[4], [5], [48], [49]

$$\rho c_p \left[\frac{\partial T}{\partial t} + (\boldsymbol{V}.\nabla)T\right] = k\nabla^2 T - \mu_0 T \frac{\partial M}{\partial T} \boldsymbol{V}.\boldsymbol{\nabla} H + \mu\Phi \qquad (6)$$

Where $\boldsymbol{v}$ denotes the velocity, $\rho$ denotes the density, $\mu$ denotes the viscosity, $\mu_0$ denotes the permeability of free space, $\boldsymbol{M}$ denotes the magnetization, $H$ denotes the magnetic field intensity, $\boldsymbol{\omega}_{\mathrm{p}}$ denotes the angular velocity of magnetic particles in the flow, $\boldsymbol{\Omega}$ denotes the vorticity in the flow, $\tau_B$ Brownian relaxation time, $I$ denotes the moment of inertia, $\tau_s$ denotes the rotational relaxation time, $\boldsymbol{M_0}$ denotes the equilibrium magnetization, $m$ denotes the magnetic moment, $n$ denotes the number of particles, $L(\xi)$ denotes the Langgevin function for paramagnetism, $k_B$ denotes the Boltzmann constant and $T$ denotes the temperature, $c_p$ denotes the specific heat at constant pressure, $k$ denotes the thermal conductivity and $\Phi$ denotes the viscous dissipation term.

Two mechanisms of ferrofluid Neel relaxation and Brownian relaxation time have an important role in the study of ferrofluid[50], [51]. This mechanism shows that the magnetization in ferrofluid can relax after changing the strength of the magnetic field[52]. In Brownian relaxation time occurs due to nanoparticles rotation of the colloidal mixture and Neel relaxation time occurs due to rotation of the magnetic vector within the particle[1], [3]–[5].

A Brownian relaxation time time $\tau_B$ is given by[1]

$$\tau_B = \frac{3\mu V}{k_B T} \qquad (7)$$

where $V = \frac{\pi(d+2s)^3}{6}$ denotes the hydrodynamic volume of the particle including surfactant layers. Here $d$ denotes the diameter of the particle and $s$ denotes the thickness of the surfactant layer.

Under certain material conditions, the magnetic moment may rotate inside the particle relative to crystal structure[39], [53]. This kind of relaxation of the colloidal particles can take place if the thermal energy is high enough to overcome the energy barrier provided by the crystallographic anisotropy of the magnetic material[4], [5]. The height of this energy barrier is given by $KV$, where $K$ is the anisotropy constant of the material[3], [5]. For the case $KV \ll k_B T$, the thermal energy is large enough to induce fluctuations of the magnetization inside the grain with a characteristic time $\tau_N$:

$$\tau_N = \frac{1}{f_0} \exp\left(\frac{KV}{k_B T}\right) \tag{8}$$

where $f_0$ is a frequency having the approximate value $10^9 \, s^{-1}$.

When $\tau_N \ll \tau_B$, relaxation occurs by the Neel mechanism, and the material is called to possess intrinsic superparamagnetism[2], [5]. When $\tau_B \ll \tau_N$, the Brownian mechanism is determined and the material exhibits extrinsic superparamagnetism[3], [5]. However, if the smaller time constant is much greater in comparison to the time scale of the experiment, then the same may be regarded as ferromagnetic[2], [53]. An effective relaxation time combined from Neel and Brown times for the relevant particle diameter can be calculated as[1], [3], [39]:

$$\tau_{eff} = \frac{\tau_B \tau_N}{(\tau_B + \tau_N)} \tag{9}$$

For the specific types of ferrofluid, the researchers are using the thermophysical properties of ferrofluid. The following mathematical equations are being used by the researchers[54]–[58]:

$$\rho_{nf} = \rho_f \left[(1-\varphi) + \varphi\left(\frac{\rho_s}{\rho_f}\right)\right] \tag{10}$$

$$(\rho c_p)_{nf} = (\rho c_p)_f \left[(1-\varphi) + \varphi \frac{(\rho c_p)_s}{(\rho c_p)_f}\right] \tag{11}$$

$$\mu_{nf} = \frac{\mu_f}{(1-\varphi)^{2.5}} \tag{12}$$

$$\frac{k_{nf}}{k_f} = \frac{k_s + 2k_f - 2\varphi(k_f - k_s)}{k_s + 2k_f + \varphi(k_f - k_s)} \tag{13}$$

Where $\rho_f$ the density of the base fluid, $\varphi$ denotes the volume concentration of nanoparticles, $(\rho_s, \rho_f)$ denotes the density of nanoparticles and base fluid, respectively, $((\rho c_p)_s, (\rho c_p)_f)$ denotes the heat capacitance of solid and base fluid, respectively, $\mu_f$ dynamic viscosity of the base fluid, $(k_s, k_f)$ denotes the thermal conductivity of nanoparticles and base fluid respectively.

### 3. Viscosity of ferrofluid

*3.1. Viscosity in the absence of the magnetic field*

In the absence of the magnetic field, the viscosity of ferrofluid depends on the volume concentration. The mathematical expression for the viscosity of ferrofluid is given as[3], [59]–[61]:

$$\mu_{(H=0)} = \mu_c \left(1 + \frac{5}{2}\tilde{\varphi}\right) \tag{14}$$

$\mu_{(H=0)}$ denotes the viscosity of ferrofluid in the absence of the magnetic fluid, $\mu_c$ denotes the viscosity of the base fluid, $\tilde{\varphi}$ denotes the volume concentration of magnetic nanoparticles including the surfactant layer.

The volume concentration $\tilde{\varphi}$ of the suspended material in the colloidal suspension can be expressed as:

$$\tilde{\varphi} = \varphi \left(\frac{d_m + 2s}{d_m}\right)^3 \tag{15}$$

Where $\varphi$ denotes the volume concentration of the magnetic nanoparticles, $d_m$ denotes the diameter of the magnetic core and $s$ denotes the thickness of the surfactant layers.

In 1970, the first improvement Eq. (14) was given as[3], [51], [62]:

$$\mu_{(H=0)} = \mu_c \left(1 + \frac{5}{2}\tilde{\varphi} + \frac{31}{5}\tilde{\varphi}^2\right) \tag{16}$$

In 1985, Rosensweig has modified the expression for the viscosity of ferrofluid as[1]:

$$\mu_{(H=0)} = \frac{\mu_c}{\left(1 - \frac{5}{2}\tilde{\varphi} + b\tilde{\varphi}^2\right)} \tag{17}$$

Where $b = \frac{\left(\frac{5}{2}\tilde{\varphi}_c - 1\right)}{\tilde{\varphi}_c^2}$ and $\tilde{\varphi}_c$ denotes the critical volume fraction of the suspended material.

Recently for the viscosity of the solution, the researchers use the following expressions[63]–[66]:

$$\mu_{H=0} = \frac{\mu_c}{(1-\varphi)^{2.5}} \tag{18}$$

*3.2. Viscosity of ferrofluid in the presence of magnetic field*

In the presence of a magnetic field, the rotation of the magnetic particle was also considered in the viscosity of ferrofluid. In 1969, researchers have introduced the theortical expressions for the viscosity of ferrofuid unde the influence of external magnetic field[67]. This expression of viscosity depends on the strength and direction of the magnetic field. This expression is[3]–[5], [68]:

$$\mu_H = \mu_s \left(1 + \frac{5}{2}\tilde{\varphi} + \frac{3}{2}\varphi' \sin^2 \varepsilon_1\right) \tag{19}$$

where $\mu_s$ denotes the viscosity of the solvent, $\varphi'$ denotes the volume fraction of the particles including the surfactant layer, $\tilde{\varphi}$ denotes the volume fraction of all suspended material including dispersants or free surfactants molecules. The term $sin^2\varepsilon_1$ includes the magnetic part in the following form[3]–[5]:

$$sin^2\varepsilon_1 = \frac{1}{2}\left(1+\frac{1}{\xi_r^2}\right) - \left[\frac{1}{4}\left(1+\frac{1}{\xi_r^2}\right)^2 - \frac{1}{\xi_r^2}sin^2\theta\right]^{\frac{1}{2}} \quad (20)$$

where $\theta$ denotes the angle between the vorticity of the flow and the magnetic field direction and $\xi_r$ denotes the ratio of the magnetic torque and the viscous torque acting on a particle. The relation between $\xi_r$ and magnetic field can be written as[4], [5]:

$$\frac{1}{\xi_r} = \frac{\mu_0 m H}{4\pi\mu_s d^3 \gamma_r} \quad (21)$$

where, $\gamma_r$ denotes the shear rate.

In a planer Couette flow, the additional viscosity due to magnetic field is given as[3], [69]–[73]:

$$\Delta\mu = \frac{\mu_0 \tau_B M_0 H}{4\left(1+\mu_0 \tau_s \tau_B M_0 \frac{H}{I}\right)} \quad (22)$$

where $\tau_s$, $\tau_B$ and $I$ are defined as:

$$\tau_s = \frac{d_1^2 \rho_1}{45\mu}, \quad \tau_B = \frac{\pi d_1^3 \mu}{2k_B T}, \quad I = \frac{1}{10}d_1^2 \rho_1 \varphi_1 \quad (23)$$

Where $d_1$ denote the mean diameter of the magnetic particles, $\rho_1$ denotes the mean density of the whole solid fraction and $\varphi_1$ denotes the volume concentration including surfactant layer. Using Eq. (20), the expression for the viscosity in Eq. (22) becomes[3], [47], [74], [75]:

$$\Delta\mu = \frac{3}{2}\varphi_1\mu\frac{\xi-\tanh\xi}{\xi+\tanh\xi} \quad (24)$$

Eq. (24) shows the expression for the rotational viscosity due to the magnetic field when the magnetic field is perpendicular to the vorticity in the flow. For arbitrary angle between magnetic field and vorticity, Eq. (24) can be written as[4], [39], [75]:

$$\Delta\mu = \frac{3}{2}\varphi_1\mu\frac{\xi-\tanh\xi}{\xi+\tanh\xi} sin^2\beta \quad (25)$$

Magnetic torque $\mathbf{M} \times \mathbf{H}$ and viscous torque $(\omega_p - \Omega)$ in ferrofluid flow generates the rotational viscosity in ferrofluids[1], [44], [73]. In the presence of the magnetic field, the fluid and particles in the colloidal suspensions rotate with different angular velocities and this difference of angular velocituies creates an additional resistance in the flow. The equilibrium of these two torques give:

$$\mu_0 \mathbf{M} \times \mathbf{H} = 6\mu\varphi_1(\boldsymbol{\omega_p} - \boldsymbol{\Omega}) \tag{26}$$

Using Eq. (26), Bacari et al. demonstrated the theoretical expression for the viscosity[39], [44], [76]

$$\Delta\mu = \frac{3}{2}\mu\varphi_1 \frac{\Omega - \omega_p}{\Omega} \tag{27}$$

The relative viscosity can be presented as[3], [42], [77]:

$$R = \frac{\mu(H) - \mu(H=0)}{\mu(H=0)} = \frac{\Delta\mu}{\mu(H=0)} \tag{28}$$

In the presence of strong the magnetic field[3]–[5]:

$$R_{(H\to\infty)} = \frac{3}{2}\varphi_1 \sin^2\beta \tag{29}$$

If the magnetic field is perpendicular to the vorticity in the flow, then the maximum relative viscosity is[3], [78]

$$R^{max} = \frac{3}{2}\varphi_1 \tag{30}$$

If size of the particle is 10 nanometers including the surfactant layer, the maximum increase in the viscosity of ferrofluid is approximately 40%. In a weak magnetic field, we consider $\tanh\xi = \xi - \frac{1}{3}\xi^3 + O(\xi^5)$, therefore,

$$\frac{\xi - \tanh\xi}{\xi + \tanh\xi} \approx \frac{1}{6}\xi^2 \tag{31}$$

A weak magnetic field represents the following expression of relative viscosity of ferrofluid:

$$R \approx \frac{1}{4}\varphi_1 \xi^2 \tag{32}$$

Researches have used these expressions of viscosity in ferrohydrodynamic flow in different regimes[79]–[84].

*3.3. Negative viscosity effects in ferrofluid*

In the presence of a stationary magnetic field, the viscosity of ferrofluid always increases. However, in the presence of an alternating magnetic field, the viscosity of ferrofluid depends not only on the strength but also on the frequency of the alternating magnetic field. In the presence of the alternating magnetic field, the relation between the angular velocity of the particle and field frequency is given by[39], [85], [86]:

$$\omega_p = \Omega\left(1 - \frac{\xi^2}{6}\frac{(1-\omega_0^2\tau_B^2)}{(1+\omega_0^2\tau_B^2)^2}\right) \qquad \text{and} \qquad \xi = \frac{\mu_0 mH}{k_B T} \tag{33}$$

Where $\omega_0$ denotes the frequency of an alternating magnetic field.

For the weak field $(mH \leq k_B T)$ the viscosity of ferrofluid is[39], [87], [88]:

$$\Delta \mu = \frac{1}{4} \mu \varphi_1 \xi^2 \frac{(1-\omega_0^2 \tau_B^2)}{(1+\omega_0^2 \tau_B^2)^2} \tag{34}$$

Here, $\omega_0 \tau_B$ denotes the dimensionless field frequency. The condition $\omega_0 \tau_B = 1$ is known as resonance condition. This condition can be achieved when the frequency of alternating magnetic fields matches with relaxation time. In this case, there is no impact of rotational viscosity due to the magnetic field[44]. A case $\omega_0 \tau_B < 1$, the expressions in Eq. (34) remains positive. This case always enhances the viscosity of ferrofluid due to applied magnetic field. For the case, $\omega_0 \tau_B > 1$, the expressions in Eq. (34) becomes negative[39], [44]. In other words, after applying a magnetic field, the viscosity of ferrofluid becomes less than without magnetic field. This viscosity reduction is known as the negative viscosity effect. If we take, $\omega_0 \tau_B \to \infty$, the impact of Eq. (34) in the viscosity of ferrofluid becomes negligible[39].

The viscosity of ferrofluid for arbitrary amplitude is[39]:

$$\Delta \mu = \frac{1}{4} \mu \varphi_1 \xi^2 \left(2 - \tanh \epsilon - 2 \frac{\tanh \epsilon}{\epsilon}\right); \epsilon = \frac{\pi}{2\omega_0 \tau_B} \tag{35}$$

Considering limit $\epsilon \to \infty$, Eq. (32) becomes $\Delta \mu = \frac{1}{4} \mu \varphi_1 \xi^2$ (viscosity due to stationary magnetic field).

## 4. Conclusions

This review on the viscosity of ferrofluid has presented the recent fundamental theoretical development of viscosity of ferrofluid. The viscosity of ferrofluid has a major role in the application of ferrofluid in sealing, biomedical engineering, and heat transfer analysis. In the presence of a magnetic field, the viscosity of ferrofluid depends on the difference between the vorticity in the flow and rotation of the magnetic nanoparticles. The magnetic field can be directed perpendicular to the vorticity in the flow to enhance the maximum viscosity. An alternating magnetic field with a higher than resonance frequency can be used to reduce the viscosity of ferrofluid. Recently, the researchers have been publishing research papers on different types of magnetic nanofluids but some of them did not consider the impact of viscosity due to applied magnetic field. For more realistic results for which theoretical and experimental results can coincide, the theoretical expressions of the ferrofluid viscosity should be considered in the computational work of magnetic fluid flow.